\documentclass[journal,transmag]{IEEEtran}
\usepackage{graphicx}
\usepackage{amsmath}
\usepackage{cite}
\usepackage{bm}

\begin{document}

\title{Effect of Pb Substitution at the Mo Site on the Magnetic Properties of the Polar Magnet Fe\textsubscript{2}Mo\textsubscript{3}O\textsubscript{8}}

\author{\IEEEauthorblockN{Takumi~Shirasaki, Taichi~Ishikawa, Shungo~Nakayama, and~Hideki~Kuwahara}
\IEEEauthorblockA{Department of Engineering and Applied Sciences, Sophia University, Tokyo 102-8554, Japan}
\thanks{Received 9 March 2026; revised 27 March 2026; accepted 30 March 2026. 
Corresponding author: H. Kuwahara (email: h-kuwaha@sophia.ac.jp)}.
\thanks{\copyright~2026 IEEE. Personal use of this material is permitted. 
Permission from IEEE must be obtained for all other uses, in any current or future media, including reprinting/republishing this material for advertising or promotional purposes, creating new collective works, for resale or redistribution to servers or lists, or reuse of any copyrighted component of this work in other works.}
\thanks{This is the accepted manuscript. 
The final published version is available at https://doi.org/10.1109/TMAG.2026.3685505}}

\IEEEtitleabstractindextext{%
\begin{abstract}
The ternary transition-metal oxide Fe$_2$Mo$_3$O$_8$ is a polar magnet characterized by a layered structure of magnetic Fe honeycomb lattices and non-magnetic Mo kagome lattices. 
Whereas previous studies have primarily focused on the chemical substitution at the Fe sites to modulate the magnetic properties, the Mo sites have remained largely unexplored due to the strong spin-singlet trimerization of Mo$^{4+}$ ions. 
In this study, we investigated the effect of substituting non-magnetic Pb$^{4+}$ and Zr$^{4+}$ ions into the Mo sites to intentionally disrupt the Mo trimers. 
Our results reveal that the disruption of the Mo spin-singlet state induces active spins within the Mo layer, resulting in the emergence of a ferromagnetic-like behavior that persists even at room temperature. 
Quantitative analysis that takes into account the weight fraction of the main phase suggests an effective spin $S = 1/2$ state per active Mo ion upon trimer disruption. 
These findings demonstrate that controlling non-magnetic cluster states within a polar host via chemical substitution is a promising approach for designing room-temperature magnetoelectric materials.
\end{abstract}

}

\maketitle
\IEEEdisplaynontitleabstractindextext
\IEEEpeerreviewmaketitle

\section{Introduction}
\begin{figure}[t]
\centering
\includegraphics[width=\columnwidth]{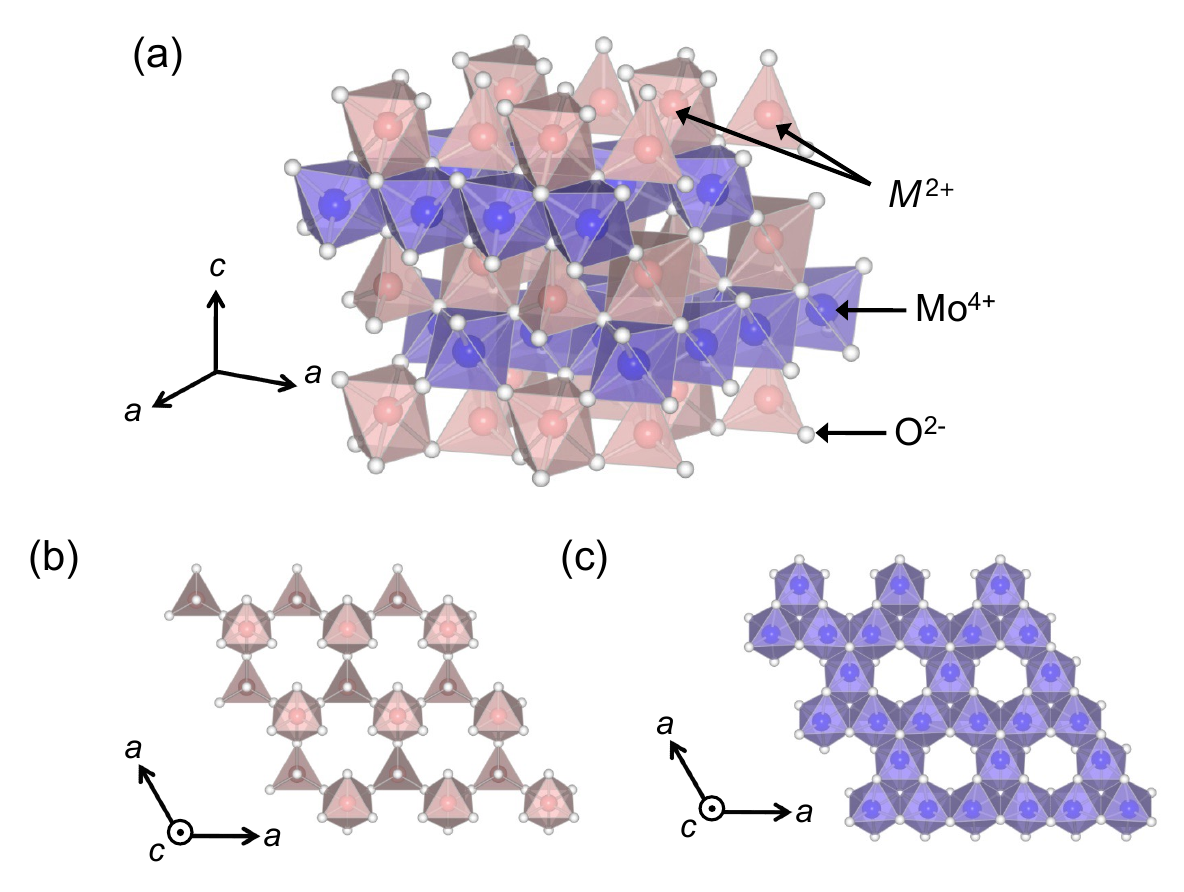}
\caption{(a) Crystal structure of the $M_2$Mo$_3$O$_8$ system (space group $P6_3mc$) visualized using VESTA\cite{Momma2011}. 
The magnetic $M^{2+}$ layers and Mo$^{4+}$ layers are alternately stacked along the $c$-axis. 
(b) Honeycomb lattice formed by $M^{2+}$ ions, occupying inequivalent octahedral and tetrahedral sites. 
(c) Kagome-like lattice formed by Mo$^{4+}$ ions, in which three adjacent Mo ions form a spin-singlet trimer, rendering the layer non-magnetic.}
\label{CrystalStructure}
\end{figure}

\IEEEPARstart{T}{he} ternary transition-metal oxides $M_2$Mo$_3$O$_8$ (where $M$ is a 3$d$ transition metal) are low-dimensional magnetic systems that crystallize in a layered structure similar to the natural mineral kamiokite\cite{McCarroll1957}\@. 
As shown in Fig. \ref{CrystalStructure}, the system belongs to the polar space group $P6_3mc$ and exhibits spontaneous electric polarization along the $c$-axis at room temperature (RT). 
The $M^{2+}$ ions occupy two distinct crystallographic sites (tetrahedral and octahedral) to form a honeycomb lattice. 
Meanwhile, the Mo$^{4+}$ ions ($4d^2$) form a kagome lattice. 
These layers alternately stack along the $c$-axis. 
Within the kagome lattice, the Mo$^{4+}$ ions form robust spin-singlet trimers, rendering the Mo layer non-magnetic\cite{Ansell1966,Abe2010}\@. 
Consequently, the bulk magnetic properties are predominantly governed by the $M^{2+}$ ions. 
The formation of these spin singlets occurs well above RT, indicating the presence of robust magnetic interactions between the Mo ions.

Extensive research has explored the magnetic properties of various $M$-site substitutions\cite{McAlister1983}\@. 
For instance, Mn$_2$Mo$_3$O$_8$ exhibits easy-plane ferrimagnetism, while Ni$_2$Mo$_3$O$_8$ shows non-collinear antiferromagnetism\cite{McAlister1984,Kurumaji2017-1,Morey2019,Yadav2023}\@. 
Among this family, Fe$_2$Mo$_3$O$_8$ has been the most actively studied. 
It undergoes an antiferromagnetic (AFM) transition at $T_{\rm N} = 60$ K with a collinear spin arrangement, where spins in the tetrahedral and octahedral sites within a single Fe layer align antiparallel\cite{McAlister1983}\@. 
This site-selective spin configuration has recently sparked renewed interest in Fe$_2$Mo$_3$O$_8$ as a prominent candidate for altermagnetism\cite{Cheong2024,Dong2025}\@. 
Furthermore, a large magnetoelectric effect associated with a spin-flip transition under a magnetic field near $T_{\rm N}$ has been reported\cite{Wang2015,Kurumaji2015,Chang2023}. 
Partial substitution of Fe by non-magnetic Zn, (Fe,Zn)$_2$Mo$_3$O$_8$, destabilizes the AFM ground state because Zn preferentially occupies the tetrahedral sites, leading to an uncompensated ferrimagnetic state\cite{Nakayama2011}\@. 
In the slightly Zn-doped regime, a variety of exotic phenomena, including electromagnons in the THz region (magnetoelectric optical effect)\cite{Kurumaji2017-2,Kurumaji2017-3,Csizi2020}, giant thermal Hall effects\cite{Ideue2017}, and magnetoelectric caloric effects\cite{Ino2025}, have been uncovered.

As described above, extensive element-substitution studies have been conducted at the $M$ sites of $M_2$Mo$_3$O$_8$. 
In this study, we turn our attention to the previously unexplored Mo layers. 
By substituting non-magnetic Pb$^{4+}$ and Zr$^{4+}$ ions at the Mo sites, we aimed to disrupt the robust Mo trimers and uncover novel functionalities within this material system. 
We report that the intentional breaking of the Mo trimers via impurity substitution successfully induces a ferromagnetic-like (FM-like) state at RT.

\section{Experimental Procedure}
Polycrystalline samples of Fe$_2$(Mo$_{1-x}A_x$)$_3$O$_8$ ($A =$ Pb, Zr; $0 \le x \le 0.20$) were synthesized using a conventional solid-state reaction. 
Stoichiometric amounts of the starting materials, Fe$_2$O$_3$, Mo, MoO$_3$, PbO$_2$, and ZrO$_2$ powders, were thoroughly wet-mixed with ethanol in an agate mortar and then pressed into pellets. 
The pellets were wrapped in platinum foil and sealed in an evacuated quartz tube. 
The samples were initially calcined at 600$^\circ$C for 20 hours and subsequently sintered at 1000$^\circ$C for 12 hours. 
This high-temperature processing yields well-crystallized bulk polycrystalline grains in the micrometer range, allowing for the evaluation of intrinsic bulk magnetic properties while minimizing influence from surface defects or uncompensated spins common in nanostructures\cite{Zhang2020}\@.

The synthesized products were ground into fine powders, and their crystal structures were characterized by powder X-ray diffraction (XRD) using Cu $K\alpha$ radiation (D8 Advance, Bruker). 
The XRD patterns were analyzed via the Rietveld refinement method using the TOPAS software\cite{Coelho2018}. 
The temperature and magnetic-field dependencies of the magnetization were measured up to 8 T using the vibrating sample magnetometer (VSM) option in a Physical Property Measurement System (PPMS-9T, Quantum Design).

\section{Results and Discussion}

\begin{figure}[t]
\centering
\includegraphics[width=0.9\columnwidth]{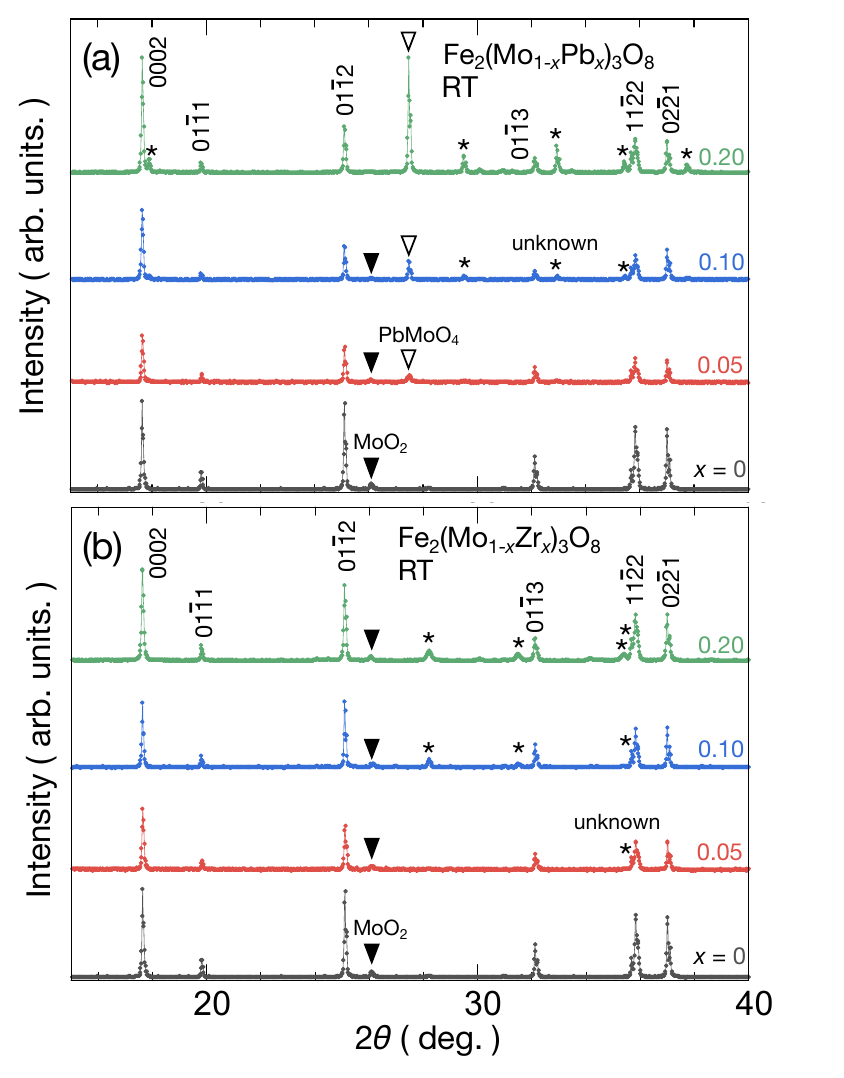}
\caption{(a) Powder X-ray diffraction patterns of polycrystalline Fe$_2$(Mo$_{1-x}$Pb$_x$)$_3$O$_8$ at room temperature (RT). 
Black and white triangles indicate peaks from MoO$_2$ and PbMoO$_4$, respectively. 
Asterisks denote peaks from unidentified impurities. 
(b) Powder X-ray diffraction patterns of polycrystalline Fe$_2$(Mo$_{1-x}$Zr$_x$)$_3$O$_8$ at RT. 
Black triangles indicate MoO$_2$ peaks, and asterisks denote unidentified impurity peaks.}
\label{XRD}
\end{figure}

Figure \ref{XRD}(a) displays the powder XRD profiles for Fe$_2$(Mo$_{1-x}$Pb$_x$)$_3$O$_8$ ($0 \le x \le 0.20$) at RT. 
For all substitution levels, the main phase was successfully indexed to the $P6_3mc$ space group, identical to the parent Fe$_2$Mo$_3$O$_8$. 
The variations in the relative intensities of the $(000\ell)$ reflections, particularly the (0002) peak, are attributed to the preferred orientation of the (0001) basal planes occurring during powder packing, a common geometric effect in layered oxides.
Traces of MoO$_2$ were detected as an impurity phase for $x = 0, 0.05,$ and $0.10$. 
In the Pb-substituted samples, the precipitation of PbMoO$_4$ was confirmed, with its fraction increasing approximately proportionally to $x$. 
It should be noted that since PbMoO$_4$ is entirely non-magnetic, its presence does not qualitatively alter the intrinsic magnetic behavior of the target material\cite{Groenink1980}\@. 
Similarly, the Zr-substituted samples were also indexed with the $P6_3mc$ space group (Fig. \ref{XRD}(b)). 
According to crystallographic databases, unlike PbMoO$_4$, stoichiometric ZrMoO$_4$ is thermodynamically unstable, and no characteristic peaks attributable to the stable Zr(MoO$_4$)$_2$ phase were detected\cite{Auray1986}\@.
Although trace amounts of unidentified impurities were observed alongside MoO$_2$, the magnetic properties of the Zr system exhibit striking similarities to those of the Pb system (as discussed below), implying the absence of FM impurity phases. 
Although both Pb$^{4+}$ (0.775 \AA) and Zr$^{4+}$ (0.72 \AA) possess larger six-coordinate ionic radii than Mo$^{4+}$ (0.65 \AA), the lattice constants did not increase monotonically with the substitution level. 
Given the significant amount of unidentified phases at $x = 0.20$, we restricted our detailed magnetic analyses to the range $x \le 0.10$, where impurity phases are relatively minimal.

\begin{figure}[t]
\centering
\includegraphics[width=\columnwidth]{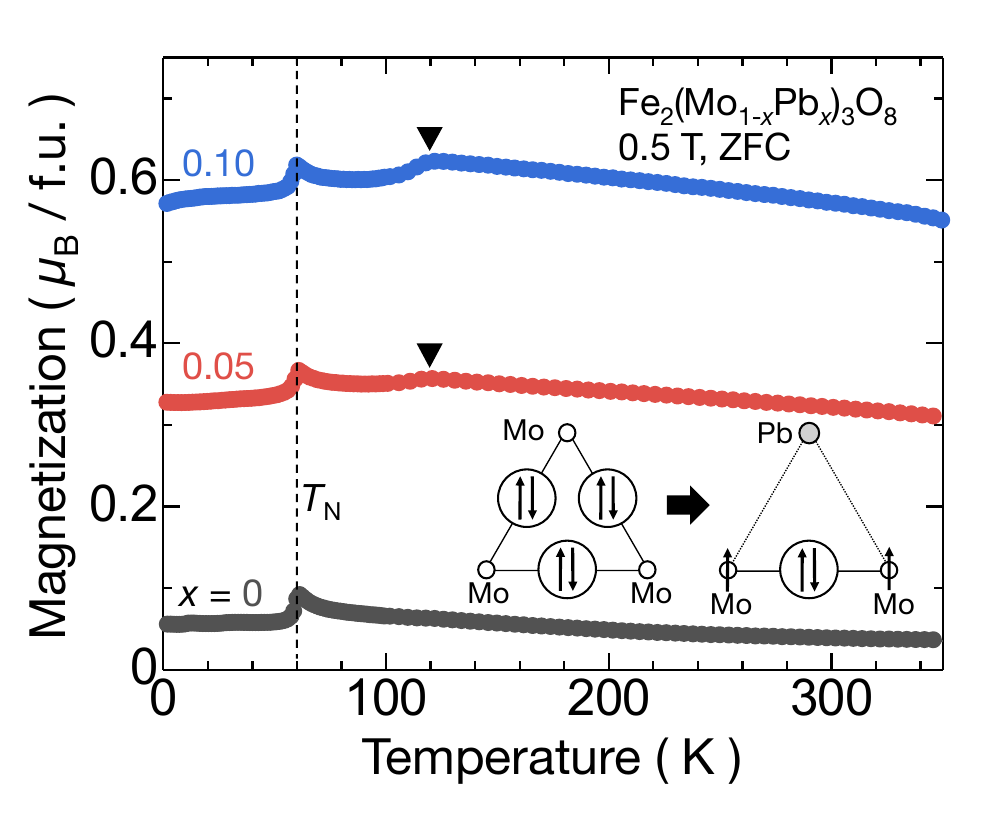}
\caption{Temperature dependence of the magnetization for polycrystalline Fe$_2$(Mo$_{1-x}$Pb$_x$)$_3$O$_8$ ($x$ = 0, 0.05, 0.10). 
The dashed line indicates the antiferromagnetic transition temperature ($T_{\rm N}$), and black triangles mark the magnetic anomalies observed for $x = 0.05$ and $0.10$. 
The inset illustrates a schematic spin configuration model that effectively explains the experimental results when a single Pb ion substitutes into a Mo trimer. 
The up and down arrows enclosed in a circle indicate a spin-singlet pair (see text for details).}
\label{PbMT}
\end{figure}

Figure \ref{PbMT} shows the temperature dependence of the magnetization for Fe$_2$(Mo$_{1-x}$Pb$_x$)$_3$O$_8$ ($0 \le x \le 0.10$) measured under zero-field-cooled (ZFC) conditions from 5 K to 350 K. 
The pristine sample ($x = 0$) exhibited an AFM transition peak at $T_{\rm N} = 60$ K, consistent with previous reports\cite{McAlister1983}. 
This transition temperature remained robust at approximately 60 K even in the Pb-substituted samples. 
However, the overall magnitude of the magnetization increased monotonically across the entire temperature range as $x$ increased. 
Furthermore, a subtle anomaly was observed near 120 K for $x = 0.05$ and $0.10$, although its detailed origin remains unclear. 
The systematic enhancement of the magnetization strongly suggests that the disruption of the Mo spin-singlet trimers by Pb substitution induces active spins within the Mo layer, as schematically illustrated in the inset of Fig. \ref{PbMT}.

\begin{figure}[t]
\centering
\includegraphics[width=\columnwidth]{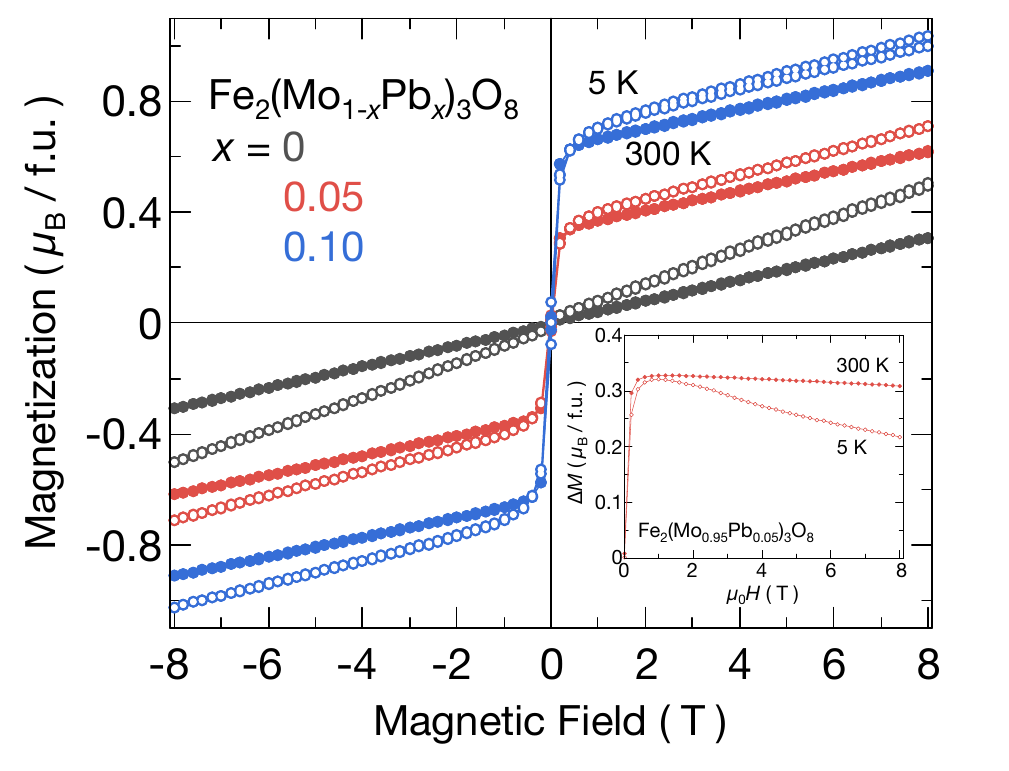}
\caption{Magnetic field dependence of the magnetization for polycrystalline Fe$_2$(Mo$_{1-x}$Pb$_x$)$_3$O$_8$ ($x$ = 0, 0.05, 0.10) at 5 K (open symbols) and 300 K (solid symbols). 
The inset shows the change in magnetization for $x = 0.05$ relative to the pristine $x = 0$ sample, $\Delta M = M(x = 0.05) - M(x = 0)$ (see text for details).}
\label{PbMH}
\end{figure}

To elucidate the variation in magnetic properties, Fig. \ref{PbMH} presents the magnetic field dependence of the magnetization at 5 K and 300 K. 
Whereas the $x = 0$ sample shows a linear field dependence typical of AFM (at 5 K) and paramagnetic (at 300 K) states, the substituted $x > 0$ samples exhibited a non-linear, FM-like increase at both temperatures. 
The magnetization at 8 T clearly increased with higher substitution levels. 
Importantly, XRD analysis confirmed the absence of any RT ferromagnetic impurities (such as Fe$_3$O$_4$). 
This implies that the uncompensated Mo spins likely exhibit a FM-like order even at RT. 
The constancy of the Fe-layer $T_{\rm N}$ suggests weak interlayer magnetic coupling between the Fe and Mo layers, indicating independent spin ordering.

To quantitatively explain this magnetic behavior, we consider the microscopic changes in the electronic states of the Mo clusters accompanying the Pb substitution. 
For the $x = 0.05$ sample, in which no unknown phases were detected, the weight fraction of the main phase was estimated to be 92 wt\% via Rietveld refinement. 
Accounting for this phase fraction, the change in magnetization from the parent compound, defined as $\Delta M = M(x = 0.05) - M(x = 0)$, is plotted in the inset of Fig. \ref{PbMH}. 
At 300 K and 8 T, $\Delta M$ is approximately 0.31 $\mu_{\rm B}$/f.u.
In the low-substitution regime such as $x = 0.05$, the probability of multiple substitutions within a single Mo-Mo-Mo trimer is extremely low and can be reasonably neglected. 
Therefore, we assume a scenario in which a trimer is disrupted by exactly one Pb ion. 
We postulate that the substitution of a single Pb ion severs two of the three original Mo-Mo spin-singlet bonds within the trimer, leaving exactly one Mo-Mo spin-singlet bond intact (see the inset of Fig. \ref{PbMT}). 
In this surviving dimer, partial singlet formation leaves each of the two Mo ions with a localized spin of $S = 1/2$. 
Assuming the spin-only $g$-factor of 2 and FM alignment of these spins under an external field, the expected saturation magnetization contributed by one disrupted trimer is 2 $\mu_{\rm B}$. 
Based on the binomial distribution $P(k) = {}_3C_k x^k (1-x)^{3-k}$, the probability of exactly one Pb ion ($k=1$) substituting into a trimer at $x=0.05$ is calculated to be approximately 13.5\%. 
The probabilities for two ($k=2, \sim 0.7\%$) and three ($k=3, \sim 0.01\%$) Pb ions occupying a single trimer are more than an order of magnitude smaller than the single-substitution case, justifying the neglect of multiple occupancy in our analysis.
Consequently, the theoretically expected saturation magnetization per formula unit is estimated as $2 \mu_{\rm B} \times 0.135 \approx 0.27 \mu_{\rm B}$/f.u. 
This theoretical value is in good agreement with the experimentally obtained magnetization increment ($\Delta M \approx 0.31 \mu_{\rm B}$/f.u.). 
These results support a scenario in which the Pb substitution breaks two of the three Mo-Mo spin-singlet bonds, thereby releasing the localized $S = 1/2$ spins at the remaining Mo sites that couple ferromagnetically.
Conversely, at 5 K and 8 T, $\Delta M$ is approximately 0.21 $\mu_{\rm B}$/f.u., which deviates slightly from the calculated value. 
We speculate that this reduction is associated with the subtle magnetic anomaly observed around 120 K. 
Independent of the Fe-layer AFM transition at 60 K, this anomaly is interpreted as an intrinsic feature of the induced Mo-layer spin system. Since the Fe spins remain paramagnetic at 120 K, it likely reflects a spin-reorientation transition of the Mo-sublattice, where the rotation of the easy axis reduces the magnetization component along the magnetic field.
While this spin-reorientation scenario serves as a highly plausible hypothesis to explain the macroscopic magnetic behavior, definitive elucidation of the precise microscopic mechanism requires future investigations using specific heat and neutron diffraction measurements.

\begin{figure}[t]
\centering
\includegraphics[width=\columnwidth]{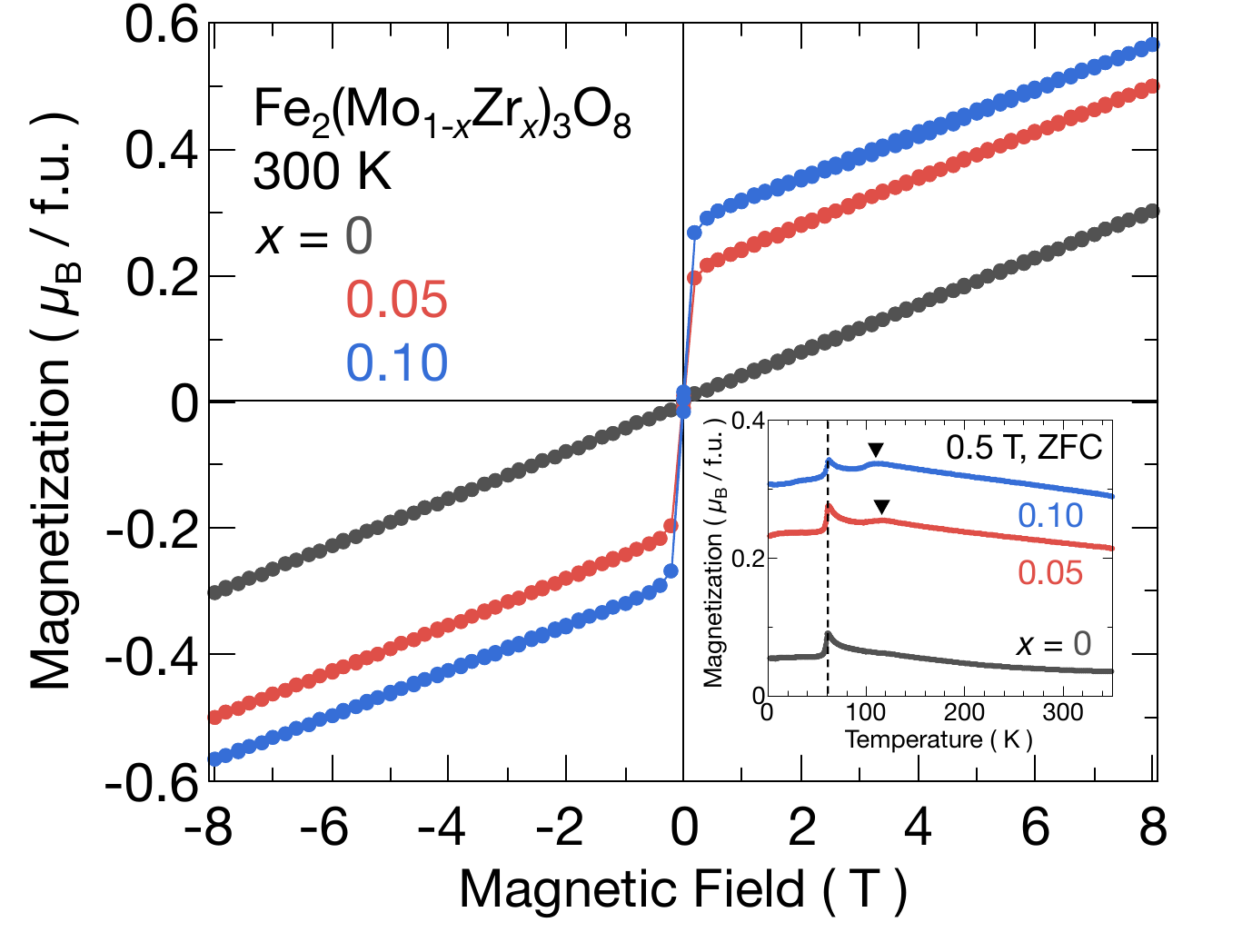}
\caption{Magnetic field dependence of the magnetization for polycrystalline Fe$_2$(Mo$_{1-x}$Zr$_x$)$_3$O$_8$ ($x$ = 0, 0.05, 0.10) at 300 K. 
The inset shows the temperature dependence of the magnetization measured under zero-field-cooled (ZFC) conditions.
The dashed line and black triangles indicate $T_{\rm N}$ and the magnetic anomalies, respectively.}
\label{ZrMTMH}
\end{figure}

Figure \ref{ZrMTMH} shows the magnetic-field dependence of the magnetization for Fe$_2$(Mo$_{1-x}$Zr$_x$)$_3$O$_8$ ($0 \le x \le 0.10$) at 300 K. 
As shown in the inset, the temperature dependence of the magnetization under ZFC conditions reveals that the AFM transition at $T_{\rm N} = 60$ K remains robust. 
Furthermore, the subtle magnetic anomaly around 120 K is clearly present, closely mirroring the behavior observed in the Pb-substituted system. 
The fact that this 120 K anomaly appears in both the Pb- and Zr-substituted samples strongly suggests that it is an intrinsic feature originating from the disruption of the Mo trimers, rather than an artifact of a specific impurity phase.
Similar to the Pb substitution, the main panel of Fig. \ref{ZrMTMH} demonstrates an increase in magnetization with increasing $x$ and a FM-like behavior at RT. 
This confirms that the emergence of FM-like behavior via Mo-site substitution is virtually independent of the specific element used, provided it effectively breaks the Mo trimer. 
The slightly smaller magnetization values at 8 T compared to the Pb case are likely attributable to an underestimation of the main phase mass due to the precipitation of unidentified impurities.

\section{Conclusion}
We investigated the modulation of magnetic properties in the polar magnet Fe$_2$Mo$_3$O$_8$ by substituting non-magnetic Pb$^{4+}$ and Zr$^{4+}$ ions into the Mo sites, where Mo$^{4+}$ ions typically form spin-singlet trimers. 
The disruption of these trimers successfully induced magnetization within the Mo layer, demonstrating FM-like behavior up to RT. 
The remarkable stability of $T_{\rm N}$ at 60 K despite the FM-like behavior in the Mo layer, indicates weak interlayer magnetic coupling. This allows the Fe and Mo layers to host distinct magnetic phases that evolve at different temperature scales, facilitating independent control of magnetic functionalities.
Given the spontaneous electric polarization inherent to Fe$_2$(Mo$_{1-x}A_x$)$_3$O$_8$, this induced FM moment opens a potential pathway to realizing RT magnetoelectric effects. 
Future work will require the synthesis of high-purity single crystals to systematically investigate the temperature evolution of the magnetoelectric coupling.
In particular, detailed dielectric and pyroelectric measurements under applied magnetic fields, together with magnetization measurements at higher temperatures, will be essential to fully elucidate the interplay between the induced magnetization and the spontaneous electric polarization.
Furthermore, impurity substitution in magnetically inert analogues such as Zn$_2$Mo$_3$O$_8$ or Mg$_2$Mo$_3$O$_8$ would allow for an isolated investigation of the Mo-layer magnetism, providing deeper insights into the physics of trimer disruption.

\section*{Acknowledgment}
The authors would like to thank D. Akahoshi for fruitful discussions. 
This work was supported in part by Sophia University Special Grant for Academic Research. 
The work of Takumi Shirasaki was supported in part by JST SPRING under JPMJSP2169 and in part by the Yazaki Memorial Foundation for Science and Technology.

\section*{Conflict of Interest Statement}
None of the authors have a conflict of interest to disclose.
\ifCLASSOPTIONcaptionsoff
  \newpage
\fi


\begin{thebibliography}{1}

\bibitem{McCarroll1957}
W. H. McCarroll, L. Katz, and R. Ward, "Some Ternary Oxides of Tetravalent Molybdenum", $J$. $Am$. $Chem$. $Soc$., vol. 79, no. 20, pp. 5410-5414, Oct. 1957\@.
\bibitem{Momma2011} 
K. Momma and F. Izumi, "VESTA 3 for three-dimensional visualization of crystal, volumetric and morphology data", $J$. $Appl$. $Cryst$., vol. 44, no. 6, pp. 1272-1276, Dec. 2011\@.
\bibitem{Ansell1966}
G. B. Ansell and L. Katz, "A Refinement of the Crystal Structure of Zinc Molybdenum(IV) Oxide, Zn$_2$Mo$_3$O$_8$", $Acta$ $Cryst$., vol. 21, no. 4, pp. 482-485, Feb. 1966\@.
\bibitem{Abe2010}
H. Abe, A. Sato, N. Tsujii, T. Furubayashi, and M. Shimoda, "Structural refinement of $T_2$Mo$_3$O$_8$ ($T$ = Mg, Co, Zn and Mn) and anomalous valence of trinuclear molybdenum clusters in Mn$_2$Mo$_3$O$_8$", $J$. $Solid$ $State$ $Chem$., vol. 183, no. 2, pp. 379-384, Feb. 2010\@.
\bibitem{McAlister1983}
S. P. McAlister and P. Strobel, "Magnetic Order In $M_2$Mo$_3$O$_8$ Single Crystals ($M$ = Mn, Fe, Co, Ni)", $J$. $Magn$. $Magn$. $Mater$., vol. 30, no. 3, pp. 340-348, Jan. 1983\@.
\bibitem{McAlister1984}
S. P. McAlister, "Unusual ferrimagnetism in Mn$_2$Mo$_3$O$_8$ and Sm$_2$In", $J$. $Appl$. $Phys$. vol. 55, no. 6, pp. 2343-2345, Mar. 1984\@.
\bibitem{Kurumaji2017-1}
T. Kurumaji, S. Ishiwata, and Y. Tokura, "Diagonal magnetoelectric susceptibility and effect of Fe doping in the polar ferrimagnet Mn$_2$Mo$_3$O$_8$", $Phys$. $Rev$. $B$, vol. 95, no. 4, Art. no. 045142 Jan. 2017\@.
\bibitem{Morey2019}
J. R. Morey, A. Scheie, J. P. Sheckelton, C. M. Brown, and T. M. McQueen, "Ni$_2$Mo$_3$O$_8$: Complex antiferromagnetic order on a honeycomb lattice", $Phys$. $Rev$. $Mater$., vol. 3, no. 1, Art. no. 014410, Jan. 2019\@.
\bibitem{Yadav2023}
P. Yadav, S. Lee, G. L. Pascut, J. Kim, M. J. Gutmann, X. Xu, B. Gao, S. W. Cheong, V. Kiryukhin, and S. Choi, "Noncollinear magnetic order, in-plane anisotropy, and magnetoelectric coupling in the pyroelectric honeycomb antiferromagnet Ni$_2$Mo$_3$O$_8$", $Phys$. $Rev$. $Research$, vol. 5, no. 3, Art. no. 033099, Oct. 2023\@.
\bibitem{Cheong2024}
S. W. Cheong and F. T. Huang, "Altermagnetism with non-collinear spins", $npj$ $Quantum$ $Mater$., vol. 9, no. 1, Art. no. 13, Jan. 2024\@.
\bibitem{Dong2025}
J. Dong, K. Wu, M. Zhu, F. Zheng, X. Li, and J. Zhang, "Nonrelativistic spin-splitting multiferroic antiferromagnets and compensated ferrimagnet with zero net magnetization", $Phys$. $Rev$. $B$, vol. 112, no. 2, Art. no. 024425, Jul. 2025\@.
\bibitem{Wang2015}
Y. Wang, G. L. Pascut, B. Gao, T. A. Tyson, K. Haule, V. Kiryukhin, and S. W. Cheong, "Unveiling hidden ferrimagnetism and giant magnetoelectricity in polar magnet Fe$_2$Mo$_3$O$_8$", $Sci$. $Rep$., vol. 5, no. 1, Art. no. 12268, Jul. 2015\@.
\bibitem{Kurumaji2015}
T. Kurumaji, S. Ishiwata, and Y. Tokura, "Doping-Tunable Ferrimagnetic Phase with Large Linear Magnetoelectric Effect in a Polar Magnet Fe$_2$Mo$_3$O$_8$", $Phys$. $Rev$. $X$, vol. 5, no. 3, Art. no. 031034, Sep. 2015\@.
\bibitem{Chang2023}
Y. Chang, Y. Weng, Y. Xie, B. You, J. Wang, L. Li, J. M. Liu, S. Dong, and C. Lu, "Colossal Linear Magnetoelectricity in Polar Magnet Fe$_2$Mo$_3$O$_8$", $Phys$. $Rev$. $Lett$., vol. 131, no. 13, Art. no. 136701, Sep. 2023\@.
\bibitem{Nakayama2011}
S. Nakayama, R. Nakamura, M. Akaki, D. Akahoshi, and H. Kuwahara, "Ferromagnetic Behavior of (Fe$_{1-y}$Zn$_y$)$_2$Mo$_3$O$_8$ (0 $\le y \le$ 1) Induced by Nonmagnetic Zn Substitution", $J$. $Phys$. $Soc$. $Jpn$., vol. 80, no. 10, Art. no. 104706, Sep. 2011\@.
\bibitem{Kurumaji2017-2}
T. Kurumaji, Y. Takahashi, J. Fujioka, R. Masuda, H. Shishikura, S. Ishiwata, and Y. Tokura, "Electromagnon resonance in a collinear spin state of the polar antiferromagnet Fe$_2$Mo$_3$O$_8$", $Phys$. $Rev$. $B$, vol. 95, no. 2, Art. no. 020405(R), Jan. 2017\@.
\bibitem{Kurumaji2017-3}
T. Kurumaji, Y. Takahashi, J. Fujioka, R. Masuda, H. Shishikura, S. Ishiwata, and Y. Tokura, "Optical Magnetoelectric Resonance in a Polar Magnet (Fe,Zn)$_2$Mo$_3$O$_8$ with Axion-Type Coupling", $Phys$. $Rev$. $Lett$., vol. 119, no. 7, Art. no. 077206, Aug. 2017\@.
\bibitem{Csizi2020}
B. Csizi, S. Reschke, A. Strini\'c, L. Prodan, V. Tsurkan, I. K\'ezsm\'arki, and J. Deisenhofer, "Magnetic and vibronic terahertz excitations in Zn-doped Fe$_2$Mo$_3$O$_8$", $Phys$. $Rev$. $B$, vol. 102, no. 17, Art. no. 174407, Nov. 2020\@.
\bibitem{Ideue2017}
T. Ideue, T. Kurumaji, S. Ishiwata, and Y. Tokura, "Giant thermal Hall effect in multiferroics", $Nat$. $Mater$., vol. 16, no. 8, pp. 797-802, May 2017\@.
\bibitem{Ino2025}
K. Ino, K. Matsuura, T. Nomoto, T. Kurumaji, Y. Tokura, and F. Kagawa, "Magnetoelectric coupling and its impact on the multicaloric effect", $Phys$. $Rev$. $B$, vol. 112, no. 2, Art. no. 024434, Jul. 2025\@.
\bibitem{Zhang2020}
L. Zhang, Y. Song, W. Wu, R. Bradley, Y. Hu, Y. Liu, and S. Guo, "Fe$_2$Mo$_3$O$_8$ nanoparticles self-assembling 3D mesoporous hollow sphere toward superior lithium storage properties", $Front$. $Chem$. $Sci$. $Eng$., vol. 15, no. 1, pp. 156-163, Nov. 2020\@.
\bibitem{Coelho2018} 
A. A. Coelho, "TOPAS and TOPAS-Academic: an optimization program integrating computer algebra and crystallographic object written in C++", $J$. $Appl$. $Cryst$., vol. 51, no. 1, pp. 210-218, Feb. 2018\@.
\bibitem{Groenink1980}
J. A. Groenink and G. Blasse, "Some New Observations on the Luminescence of PbMoO$_4$ and PbWO$_4$", $J$. $Solid$ $State$ $Chem$., vol. 32, no. 1, pp. 9-20, Mar. 1980\@. 
\bibitem{Auray1986}
M. Auray, M. Quarton, and P. Tarte, "New Structure of High-Temperature Zirconium Molybdate", $Acta$ $Cryst$., vol. C42, no. 3, pp. 257-259, Mar. 1986\@.

\end{thebibliography}
\end{document}